\documentclass[%
 reprint,
 amsmath,amssymb,
 aps,
]{revtex4}

\usepackage{graphicx}
\usepackage{dcolumn}
\usepackage{bm}
\usepackage{xcolor}
\usepackage{tikz}
\usepackage{caption}
\usepackage{subcaption}

\usepackage{xcolor}

\begin{document}

\preprint{APS/123-QED}

\title{Stable Attractors for Neural networks classification via Ordinary Differential Equations (SA-nODE)}

\author{Raffaele Marino}
 \altaffiliation{Dipartimento di Fisica e Astronomia, Università degli studi di Firenze.}
 \email{raffaele.marino@unifi.it}
\author{Lorenzo Buffoni}
\altaffiliation{Dipartimento di Fisica e Astronomia, Università degli studi di Firenze.}
\author{Lorenzo Chicchi}
\altaffiliation{Dipartimento di Fisica e Astronomia, Università degli studi di Firenze.}
\author{Lorenzo Giambagli}
\altaffiliation{Dipartimento di Fisica e Astronomia, Università degli studi di Firenze.}
\author{Duccio Fanelli}
\altaffiliation{Dipartimento di Fisica e Astronomia, Università degli studi di Firenze.}
\date{\today}

\begin{abstract}
A novel approach for supervised classification is presented which sits at the intersection of machine learning and dynamical systems theory. At variance with other methodologies that employ ordinary differential equations for classification purposes, the untrained model is a priori constructed to accommodate for a set of pre-assigned stationary stable attractors. Classifying amounts to steer the dynamics towards one of the planted attractors, depending on the specificity of the processed item supplied as an input. Asymptotically the system will hence converge on a specific point of the explored multi-dimensional space, flagging the category of the object to be eventually classified. Working in this context, the inherent ability to perform classification, as acquired ex post by the trained model, is ultimately reflected in the shaped basin of attractions associated to each of the target stable attractors. The performance of the proposed method is here challenged against simple toy models crafted for the purpose, as well as by resorting to well established reference standards. 
Although this method does not reach the performance of state-of-the-art deep learning algorithms, it illustrates that continuous dynamical systems with closed analytical interaction terms  can serve as high-performance classifiers. 
\end{abstract}

\maketitle


\section{Introduction}\label{sec:intro}

Machine learning (ML) \cite{bishop2006pattern, shalev2014understanding} and deep learning (DL) \cite{lecun2015deep, goodfellow2016deep, prince2023understanding} stand at the forefront of global technology use and research \cite{liu2023summary}. In an age where data complexity is ever-increasing, these tools excel in uncovering novel patterns, thereby enabling the achievement of groundbreaking results. Their profound impacts are indeed felt across myriad of disciplines, from medicine to optimization, partial differential equations and finance  \cite{esteva2019guide, ching2018opportunities, razzak2018deep, marino2023solving, heaton2017deep, sezer2020financial, marino2021learning}. The escalating sophistication of ML and DL technologies poses pressing challenges, chiefly in the realm of interpretability \cite{rudin2019stop}, pointing to the need for novel technical solutions that could 
contribute to elucidate the process that underlies decision making, which is often obscured from our understanding. 

In standard approaches to dense neural network training one seeks to optimise the weights that link pairs of neurons associated to adjacent layers. This is achieved by computing the gradient of the imposed loss with respect to the weights. This procedure amounts to operate in the so called direct space. Alternatively, the learning can be reformulated in reciprocal space, the spectral attributes (eigenvalues and eigenvectors) of the underlying transfer operators being the actual target of the optimisation. This procedure introduced in \cite{giambagli2021machine} allows for a substantial compression of the space of trainable parameters. Eigenvalues do provide a reliable ranking of the nodes, an observation which can be used, downstream of training, to prune unessential computing units and return a significantly compressed network, with almost identical classification abilities \cite{buffoni2022spectral}. Neural networks, in their original conception, are static entities. As such, they do not incorporate time as a distinctive algorithmic element. 

To fill this gap, Neural Ordinary Differential Equations (Neural ODEs) \cite{chen2018neural, bishop2023deep} have been proposed as a prime example of the intersection between dense neural networks and dynamical systems. Neural ODEs illustrate that the information flow within a neural network can be seen as the evolution of a system over time. Unlike traditional networks that have a fixed number of hidden layers, these models parametrize the derivative of the hidden state, using a neural network to determine the layer's depth dynamically. The model's output is calculated using a differential equation solver, allowing it to adapt to different inputs and balance numerical precision with computational speed. The training process is enhanced by a scalable method for backpropagation through any ordinary differential equation (ODE) solver that implements a strategy known as Adjoint Method \cite{pontryagin2018mathematical}. A correspondence can be drawn between Neural ODEs and Skip – Connection Neural Networks \cite{He2016resnet, he2016deep} where the skip connection acts from one layer to the following one. Notably, the weights should be the same for every step for the equivalence to hold. Relevant is also the setting of Liquid Time-Constant Networks \cite{hasani2021liquid} which implement the biologically inspired idea of an intrinsic relaxation time of activation and continuous time recurrent neural network \cite{FUNAHASHI1993801, Randall1995, 6814892}.

Working in this setting, we here propose a variant of the Neural ODEs which accommodates for an analytical expression of the 
computing non linear kernel. Further, and by leveraging on the spectral formulation of the coupling operator,  the untrained model is a priori furnished with a set of stationary stable attractors. Classifying implies shaping the basin of the planted attractors, in such a way that items belonging to different categories, supplied as an input, will be eventually directed towards distinct targets.  These latter will uniquely flag the class of pertinence of the object to be eventually classified. The ability of the system to cope with a given classification task will be therefore inherited by the characteristics of the basin of attractions, as sculpted upon training. In our view, this paves the way for a better conceptual grasping of the overall decision making process, in 
terms of a genuine dynamical system that flows in time, by funnelling different input towards their deputed destination site. 

More concretely, we will focus on image classification tasks. The network which defines the backbone of the examined dynamical model is made of $N$ nodes, $N$ denoting the number of pixels of the images to be classified. Each node is assigned with a double well potential and decorated with a scalar and continuous state variable which can explore the landscape of the assumed potential. Nodes are then sensing its nearest neighbours, as stipulated by a linear coupling term. The web of inter-nodes connections is stored in a   
$N \times N$ adjacency weighted matrix, whose elements represent the target of the training process. The above matrix is formulated in reciprocal domain: a subset of suitably tailored eigenvectors is assigned to its kernel and define the attractive poles for the globally coupled dynamics. The eigenvalues populate a specific interval of the real axis that yields stability of the crafted attractors, as dictated by a linear stability analysis. In essence the model to be trained is an extended collection of interacting units subject to contrasting tendencies: local reactions force the system towards the minima of the potential, while global, spatially (across nodes) extended, interactions make the system to evolve towards complex heterogeneous stable states. The algorithmic scheme proposed will be denoted in short SA-nODE, Stable Attractors for Neural Networks Classification via Ordinary Differential Equations.

After having established the theoretical background for the proposed methodology, we will turn to presenting an extensive collection of tests. These include exploring the impact of random perturbations on the algorithm's performance.  The paper is organized as follows: 
In Sec. \ref{sec::model}, we will formulate the model, including the strategy to embed stable attractors. Sec. \ref{sec:train} details our experimental setup and explains the training approach employed. In Sec. \ref{sec:letterdata}, we present results from a synthetic dataset composed of letters of size $7\times 7$ \cite{lecun1998gradient}. Sec. \ref{sec:MNIST} demonstrates the performance of the algorithm on two well-known benchmark datasets, namely MNIST and Fashion MNIST. In Sec. \ref{sec:discuss}, we interpret the implications of our results and conclude our manuscript.

\section{The model}
\label{sec::model}

In this section, we will introduce the basic principles that underlie the functioning of SA-nODE. As anticipated, the proposed classification model is a learnable autonomous dynamical system, composed of $N$ interacting neurons, which can be cast in the general form: 

\begin{equation} \label{eq:model}
    \vec{\dot{x}}(t)=\vec{F}(\vec{x}),  
\end{equation}

where $ \vec{x}\in \mathbb{R}^N $, and $\vec{F}(\vec{x})=-\vec{\nabla}V(\vec{x}(t), a, \gamma)+\beta\mathbf{A}\vec{x}(t)$. The vector  $ \vec{x}$ has entries of $O(1)$, while $ \vec{\dot{x}}(t) $ represents the derivative with respect to time of $\vec{x}(t)$. $ V(\vec{x}(t), a, \gamma):\mathbb{R}^N \to \mathbb{R} $ is a scalar field, i.e., a potential; with $ \vec{\nabla} $ we define the gradient operator. The adjacency matrix describing the graph is $ \mathbf{A} \in \mathbb{R}^{N \times N} $. Such a matrix encapsulates information on existing (weighted and sign sensitive) interactions between the nodes of the network. $\beta$, $\gamma$ and $a$ are parameters of the system.

\subsection{Planting the attractors.}
\label{subsec::model}

As anticipated, we postulate a double well potential, expressed as $V(\vec{x}(t), a, \gamma)=\gamma(\vec{x}^2(t) - a^2\vec{\mathbf{1}})^T(\vec{x}^2(t) - a^2\vec{\mathbf{1}})$. In this formulation, each element of $\vec{x}^2$ is denoted as $x^2_i$, $i$ ranging from $1$ to $N$. In the uncoupled limit ($\beta=0$), each node asymptotically approaches one of the two available equilibria, respectively locates in $\pm a$. The uncoupled model displays hence $2^{N}$ distinct attractors, i.e. the whole set of independent combinations obtained by permuting the two above equilibria across the collection of $N$ available nodes. These latter are no longer solutions of the examined model when $\beta$ is made different from zero. In the following, we choose $\beta = \frac{1}{\sqrt{N}}$. 
This scaling is in line with the model proposed in \cite{sherrington1975solvable, panchenko2013sherrington}, where the authors assume that the entries of $\mathbf{A}$ are independently and identically distributed as $O(1)$ variables from a standard Gaussian distribution. The value  of $\gamma$, which sets the strength of the potential, will be self-consistently adjusted through training. It should be emphasized that a prototypical double well potential is found to emerge in models relevant to computational neuroscience, as follow the intertwined interaction between distinct families of excitatory and inhibitory neurons \cite{zankoc2017diffusion}. In this respect, the simplified model that we have here formulated unlocks a perspective view full of captivating biomimetic implications.

As previously indicated, the matrix $\mathbf{A}$ is a critical component as it incorporates the structural details of the network's topology. Assume $\mathbf{A}$ to be represented as $\mathbf{A}=\Phi \Lambda \Phi^{-1}$, where $\Phi$ belongs to the set of real matrices $\mathbb{R}^{N \times N}$ and $\Phi^{-1}$ signifies its inverse. Within this framework, the matrix $\Phi$ comprises the eigenvectors of $\mathbf{A}$, arranged as its columns. Meanwhile, the matrix $\Lambda$, also in $\mathbb{R}^{N \times N}$, is a diagonal matrix containing $\mathbf{A}$'s eigenvalues. For ease of reference, we denote the columns of $\Phi$ as $\vec{\phi}^{(l)}$, where $l$ ranges from 1 to $N$. This approach of decomposing the interaction matrix mirrors the spectral approach to machine learning, as outlined in references \cite{giambagli2021machine, buffoni2022spectral, chicchi2021training, chicchi2023recurrent}. The whole idea of SA-nODE is to a priori plant a congruous number of non linear stable attractors within the full untrained model. Indeed the $ k=1,\dots,K $ columns of matrix $ \Phi $, where $ K $ stands for the total number of classes to be eventually categorised, correspond to our identified attractors. To reach our goal we assign the entries of the selected vectors $ \vec{\overline{\phi}}^{(k)} $ to take values  $ \pm a $, namely the positions of the fixed points of the a-spatial dynamics ($\beta=0$). In formulae, we require $\left(\vec{\overline{\phi}}^{(k)}\right)_i=\pm a $, $\forall i=1,...,N$. Further, we place the target vectors $ \vec{\overline{\phi}}^{(k)} $ in the kernel of $\mathbf{A}$. In other words, the matrix $ \Lambda $ has eigenvalues $ \lambda^{(k)}=0 $, $\forall k=1,...,K$, for satisfying the property that the corresponding eigenvectors sits in the kernel of $\mathbf{A}$. Thus, in light of the above prescriptions, $\vec{x}^k_{st} = \vec{\overline{\phi}}^{(k)}$ $ k=1,\dots,K $ are stationary solutions ($\dot{\vec{x}}=0$) of the examined model: when the solution aligns to $\vec{x}^k_{st}$ the linear inter-nodes coupling term disappear, by construction. Further the individual components of $\vec{x}^k_{st}$ are chosen in such a way $-\vec{\nabla}V(\vec{x}^k_{st}, a, \gamma)=0$, thus yielding the sought condition $\dot{\vec{x}}=\vec{F}(\vec{x}_{st})=0$.

However, merely planting attractors is insufficient. Our objective is to enforce asymptotic stability, following the procedure detailed above.
 
\subsection{Enforcing linear stability}
\label{subsec::linearstability}

To investigate the linear stability, we introduce a perturbation $\delta \vec{x}$ around $\vec{\overline{\phi}}^{(k)}$. Each component of the perturbation is defined as follow: $\delta x_i=x_i-\overline{\phi}_i^{(k)}$, with $\delta{x_i}$ representing a small deviation from the state $\overline{\phi}_i^{(k)}$. The model, i.e., equation \eqref{eq:model}, is then linearized around $\vec{\overline{\phi}}^{(k)}$, leading to the following expression for each component:

\begin{equation}\label{eq::picosc}
\begin{split}
   &\delta \dot{x}_i=f(\overline{\phi}_i^{(k)})+ f'(\overline{\phi}_i^{(k)}) \delta x_i + \\
   &\beta \sum_{j=1}^N A_{ij}\overline{\phi}_j^{(k)} +\beta\sum_{j=1}^N A_{ij}\delta x_j +O(\delta x_i^2),
\end{split}
\end{equation}
where $f(\overline{\phi}_i^{(k)})=-\frac{\partial V(\vec{x}, a, \gamma )}{\partial x_i}\bigg |_{x_i=\overline{\phi}_i^{(k)}}$ and   $f'(\overline{\phi}_i^{(k)})$ identifies the derivative with respect to the variable $x_i$.
Higher-order terms in $\delta x_i$ are neglected in this treatment. In \eqref{eq::picosc}, the term $f(\overline{\phi}_i^{(k)})=f(\pm a)=0$ by construction, as well as $\beta \mathbf{A}\vec{\overline{\phi}}^{(k)}=\vec{0}$. Equation \eqref{eq::picosc} can, therefore, be further simplified to:

\begin{equation}\label{eq::picoscfinale}
\delta \dot{x}_i=f'(\pm a) \delta x_i  +\beta\sum_{j=1}^N A_{ij}\delta x_j
\end{equation}

To proceed with the linear stability analysis, we aim to express $\delta x_i$ in the basis of eigenvectors of $\mathbf{A}$. In this context, $\delta x_i$ with $\sum_{\alpha}c_\alpha \phi_i^{(\alpha)}$, with 
$\mathbf{A}\vec{\overline{\phi}}^{(\alpha)}=\mathbf{\Lambda}\vec{\overline{\phi}}^{(\alpha)}$, where, we recall, $\mathbf{\Lambda}$ is diagonal.

By substituting $\delta x_i$ with the expression above, we obtain an equation for the evolution of coefficients $c_\alpha$. This equation is given by:

\begin{equation}\label{eq::evolcoeff}
\sum_{\alpha=1}^N\phi_i^{(\alpha)}\left\{\dot{c}_\alpha-\left( f'(\pm a) +\beta \lambda^{(\alpha)} \right)c_\alpha \right\}=0,\,\, \forall \alpha.
\end{equation}

The condition for a stable solution of the differential equation $\dot{c}_\alpha=( f'(\pm a) +\beta \lambda^{(\alpha)})c_\alpha$ is thus provided by $\lambda^{(\alpha)}<\frac{8 a^2 \gamma}{\beta}$, $\forall \alpha$. From the solution of the differential equation, we can immediately observe that if $\lambda^{(\alpha)}<\frac{8 a^2 \gamma}{\beta}$, $\forall \alpha$, then $\vec{\overline{\phi}}^{(k)}$ is a stable attractor of the dynamics, meaning that perturbations decay exponentially with time around it. Conversely, if there exists at least one $\lambda^{(\alpha)}>\frac{8 a^2 \gamma}{\beta}$ then  the attractor $\vec{\overline{\phi}}^{(k)}$ becomes unstable, as a small perturbation grows exponentially with time. Moreover, in the case of stability, we can define $t_s=\frac{1}{\min |f'(\pm a) + \beta \lambda|}$, which indicates the characteristic time scale for an imposed perturbation to vanish.

In conclusion, we have examined the linear stability of the imposed attractors, the eigenvectors of the coupling matrix assigned to its kernel.
The analysis yields an upper bound for the set of eigenvalues $\lambda^{(\alpha)}$, a condition that should be enforced for the attractors to prove stable to external perturbations.

\section{Training the model and the metric of convergence}\label{sec:train}

Following the analysis carried out above, we are now in a position to introduce the training process that the algorithm must undergo to learn to classify images. Without loss of generality, we fix the parameters $a$ at $0.5$ and let $\gamma$ to be trainable. 
From the observations made in the previous section, we constrain the eigenvalues of the matrix during the learning process to take any value in the range $(- \infty, \frac{8a^2\gamma}{\beta})$. As an initial condition for $\gamma$ we opt for a value equal to $0.4$, while  trainable eigenvalues are initialized as $\mathcal{N}(-5, 1)$, i.e. a normal distribution with mean equal to $-5$ and variance equal to $1$.

With this stated, we construct a statistical learning model where only the components of the matrix $\Phi$, which are not the embedded eigenvectors, as well as the non-zero eigenvalues can be self-consistently learned. In other words, the first $K$ columns of matrix $\Phi$, those identifying the embedded attractors, are not to be learned, just as the first $K$ eigenvalues of the matrix  $\Lambda$, which are fixed to be zero by construction. As an initial condition for the non-embedded eigenvectors, we opt for orthogonal random eigenvectors.

\begin{figure}[htbp]
    \includegraphics[width=\textwidth]{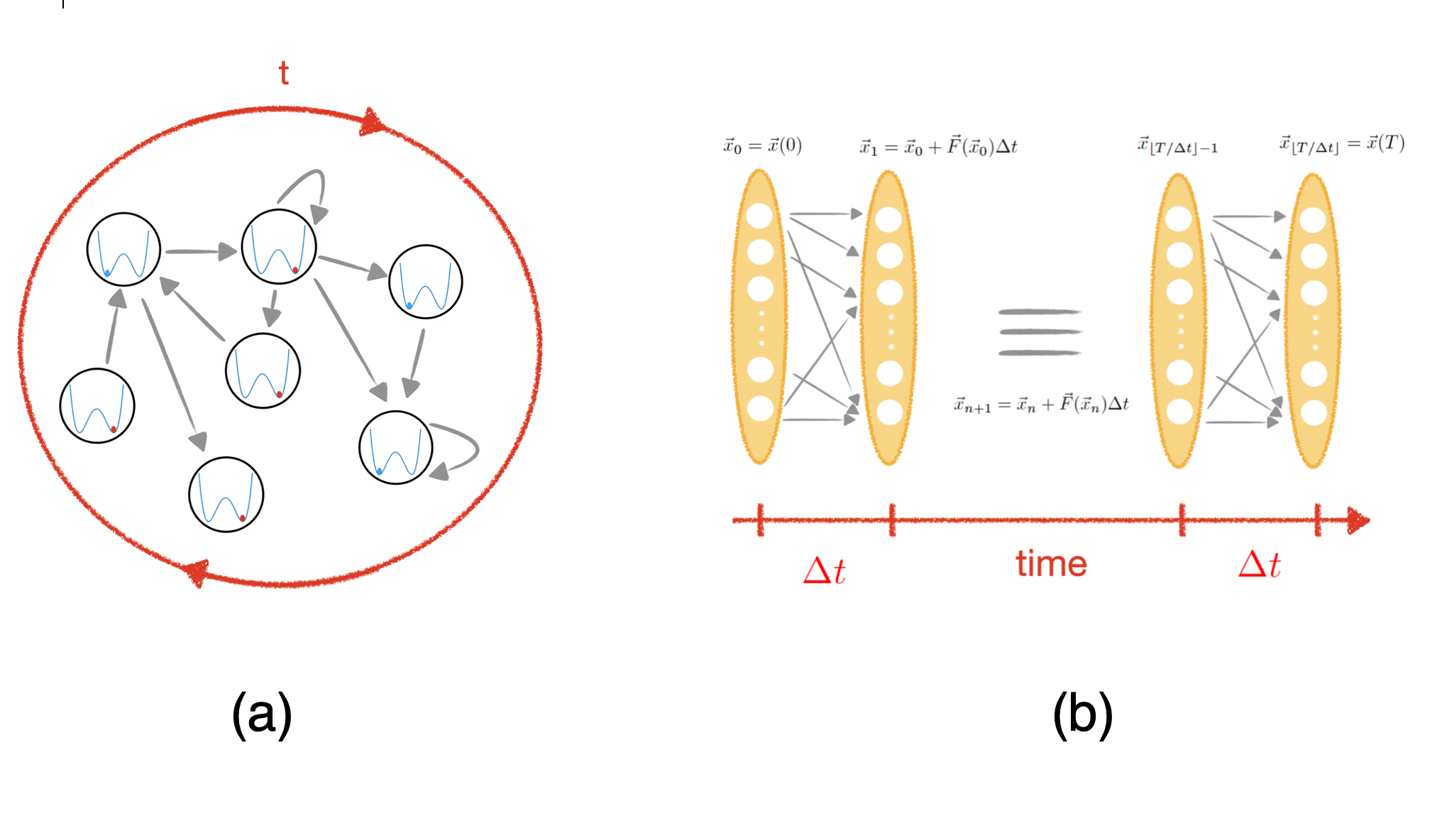} \\
    \caption{\textbf{Panel (a)}: Schematic representation of the dynamical model employed. Each neuron is uniquely associated with a single pixel of the image to be classified. The local dynamics is driven by a double well potential, as pictorially depicted.  \textbf{Panel (b)}: Schematic representation of the discrete Euler version of the examined continuous dynamical model, implemented as a recurrent neural network.     
    }
    \label{fig1NN}
\end{figure}

As our system evolves over time, the values of $\vec{x}$ are updated. The updating is performed by an Euler algorithm implemented as a recurrent neural network, following the scheme depicted in Fig. \ref{fig1NN}. Identical layers made of $N$ nodes are mutually linked by a linear coupling matrix $\mathbf{A}$, computed via its spectral decomposition, which contains the parameters to be trained. Non linear filters acting on each node follow the interaction term here postulated, i.e. the symmetric double well potential. Time flows along the horizontal axis, two successive layers of the imposed feedforward deep architecture being separated by a finite amount, $\Delta t << 1$. By operating in this framework, enables us to resort to standard optimization tools, as routinely employed within the machine learning community, to train the proposed dynamical model. By choosing $\Delta t$ sufficiently small we make sure that the trained discrete model behaves like its continuous counterpart. In concrete terms, we therefore train an ensemble made of continuous and linearly coupled ordinary differential equations, which display a set of assigned asymptotic attractors, to operate as genuine classifier. In the annexed Appendix \ref{app::continuity}, we consider a variant of the integration method that makes use of a Runge-Kutta scheme. The underlying idea is identical to what previously illustrated, except for the fact that for the deployment on an equivalent recurrent neural network one needs to accommodate for multiple evaluations of the updated state (namely account for additional layers), as following the prescription of the employed numerical algorithm. Tests on the convergence of the algorithm are enclosed in the aforementioned Appendix \ref{app::continuity}.

The parameter space for optimization resides in $ \mathbb{R}^{(\mathcal{N}+1)(\mathcal{N}-K) + 1} $, where $ K $, we recall, denotes the number of classes in our classification problem, and the singular dimension relates to the parameter $\gamma$. In practice, we take a dataset $ \mathcal{D} = (\vec{x}, y)^{(j) \in [1, \dots, |\mathcal{D}|]} $ of size $ |\mathcal{D}| $, where $ \vec{x}^{(j)} $ is an input datum, and $y^{(j)} $ represents the target, i.e., a number that ranges from 1 to $K$. This target $y$ is mapped into our attractor $ \vec{\overline{\phi}}^{(y^{(j)})} $ of the dynamics. In other words, the target $y=1$ is the first eigenvector $\vec{\overline{\phi}}^{(1)}$, the target $y=2$ is the second eigenvector $\vec{\overline{\phi}}^{(2)}$ and so on. 
Each analyzed dataset is split into a training and a test sets.

The objective of the training is to minimize the loss function $ \mathcal{L}=\frac{1}{|\mathcal{D}| }\sum_{j=1}^{|\mathcal{D}| } (\vec{x}^{*{(j)}}-\vec{\overline{\phi}}^{(y^{(j)})})^T(\vec{x}^{*{(j)}}-\vec{\overline{\phi}}^{(y^{(j)})}) $.  
$\vec{x}^*$ is the value of $\vec{x}^*=\vec{x}(T)$, when $T$ is large enough, i.e. the dynamical system in \eqref{eq:model} reaches its stationary state. In other words, we give as an initial condition to our dynamical system the input datum, i.e., $\vec{x}(0)=\vec{x}^{(j)}$, and let it evolve for a time $T$ large enough. As anticipated, and in the minimal scheme here discussed,
the evolution of the system is performed using Euler' algorithm, i.e. $\vec{x}_{n+1}=\vec{x}_n+\vec{F}(\vec{x})\Delta t$, where $\Delta t$ is the integration step and $n$ is the index for the $n$-th iteration. The maximum number of iterations is fixed a priori. Given a time $T$, and chosen the value for $\Delta t$, the total number of iterations is defined as $n_{max}=\lfloor \frac{T}{\Delta t}\rfloor$. 
Reached the time $T$, the value of $\vec{x}^{*(j)}=\vec{x}^{(j)}(T)$ is used for optimizing the loss function.  The optimization process can be facilitated using  Adam algorithm \cite{kingma2014adam} (with default learning rate).

Once optimal weight configuration is attained via Adam, we can proceed to evaluate the dynamical system's performance for classifications on test set elements. Importantly, these test set elements are a subset of the dataset $ \mathcal{D} $ that the system has never encountered during the training process. The number of epochs for each analysis in this manuscript is fixed at $700$ and mini-batch size equal to $250$. The hyper-parameters were systematically optimized to achieve the best performance. The code for reproducing the result can be downloaded from \cite{MarinoGithSANODE2024}.

To evaluate the effectiveness of our algorithm, we must compute a metric defining the system's performance. We opt to compare the output images with their corresponding targets. To this end, we   compute the overlap between the final state and the corresponding target. Mathematically, we define such overlap as:
\begin{equation} \label{magnetization}
\begin{split}
    &m_f=\frac{1}{|\mathcal{D}_{test}|}\sum_{j=1}^{|\mathcal{D}_{test}|}\frac{(\vec{x}^{*(j)})^T\vec{\overline{\phi}}^{(y^{(j)})}}{Na^2}=\\
    &=\frac{1}{|\mathcal{D}_{test}|}\sum_{j=1}^{|\mathcal{D}_{test}|}\frac{1}{Na^2}\sum_{i=1}^N x^{*(j)}_i\overline{\phi}^{(y^{(j)})}_i,
\end{split}
\end{equation}
where $|\mathcal{D}_{test}|$ is the cardinality of the test set. $m_f$ can be seen as a sort of averaged magnetization of the system, as well described in disordered systems \cite{mezard1987spin}.

As alternative measure of the convergence to the deputed attractors, we introduce the Mean Squared Error (MSE), which can be monitored over time $t$. The MSE is defined as:
\begin{equation}\label{eq:MSE}
MSE(t) = \frac{1}{B} \sum_{m=1}^{B}(\vec{x}(t)^{(m)} - \vec{\overline{\phi}}^{(y^{(m)})})^T(\vec{x}(t)^{(m)} - \vec{\overline{\phi}}^{(y^{(m)})}).
\end{equation}
where $m$ stands for the $m$-th instance, and $B=|\mathcal{D}_{sample}|$ denotes the sample size. Remark that $\vec{x}(t)$ represents the time evolution of the trained model, for any given item supplied as an input.   

To enhance the complexity of the simple datasets used in our preliminary evaluation (see below for a discussion on this issue)
and to mimic real-world scenarios where data may be subject to noise, we introduce uniform noise to individual pixels, chosen randomly. This alteration effectively \textit{dirties} the image to be processed, offering a more realistic portrayal of what might be encountered in practical applications. By adjusting the number of pixels to be corrupted with $\epsilon \in [0,1]$, the level of noise can be controlled, thereby allowing for systematic studies of noise resilience in the proposed model. In other words, the number of pixels corrupted, $\zeta$, is $\zeta=\epsilon\, N$, where $N$ is the size of the system, i.e., the number of total pixels of an image. 

This deliberate induction of noise will be inserted into the training set and/or the test set. In such cases, we use $\epsilon_{train}$ to define the level of noise injected into the training set, which means that the model is trained with images that are corrupted with a particular level of noise, equal to $\epsilon_{train}$. Similarly, we use $\epsilon_{test}$ to define the level of noise injected into the test set, once a particular model has been trained with its own level of noise.

\section{Letter Dataset}\label{sec:letterdata}

\begin{figure*}
    \centering
    \includegraphics[width=\textwidth]{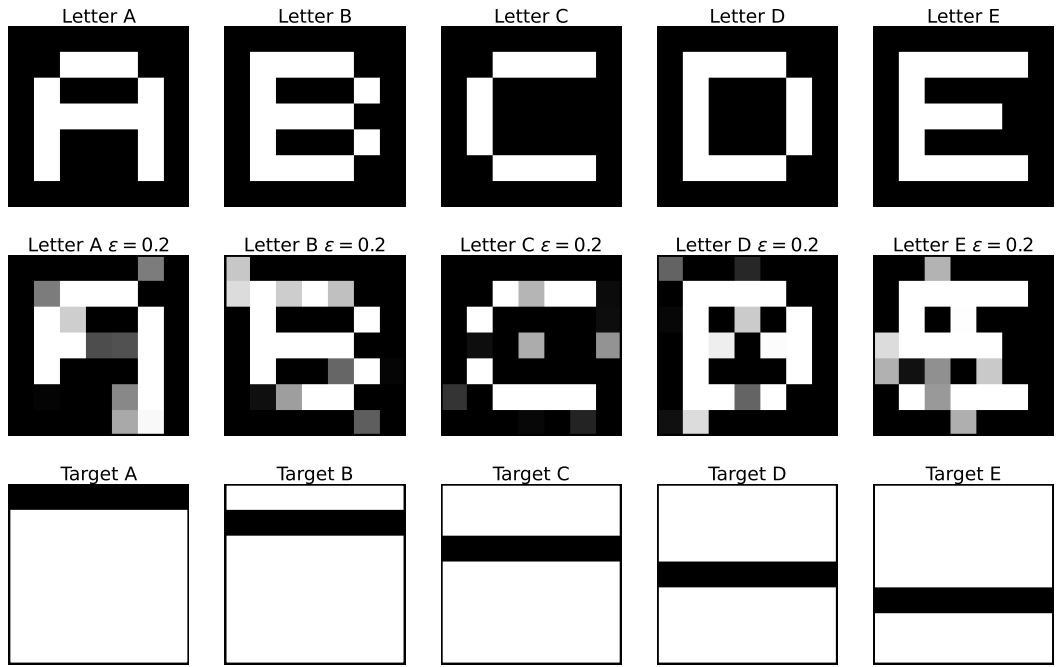}
    \caption{Representation and corresponding targets (asymptotic attractors) of the letters 'A', 'B', 'C', 'D', and 'E' as $7 \times 7$ gray scale. The top row illustrates the representation of each letter, enclosed within rectangular borders.  The middle row displays the same letters but with a noise factor of $\epsilon=0.2$ applied, introducing slight distortions. The bottom row presents the target pattern for each letter, with the black line of the target for 'B' starting immediately after the line of the target for 'A', the target for 'C' starting after 'B', 'D' after 'C', and 'E' after 'D', each surrounded by a sea of white pixels. This configuration allows for precise mapping to each corresponding letter, serving as the planted attractors within the matrix $\Phi$, as illustrated in the main body of the paper. Recall in particular that the attractors are shaped by employing the two entry values $\pm a$. Here, $-a$ refers to pixel colored in white, black pixels are associated to $a$.}
   \label{inputtarget5}
\end{figure*}

In this section, we present a synthetic dataset constructed through binarized ASCII characters, specifically tailored to our experimental requirements. Specifically, the chosen 
set involves five letters, from 'A' to 'E', each formulated as $7 \times 7$ in gray scale, see Fig. \ref{inputtarget5}. This synthetic dataset provides a valuable resource for validating our algorithm's performance under various noise conditions, introduced as follows the procedure highlighted above. At this stage the objective is not to benchmark the performance of our learning model against conventional metrics of accuracy and efficiency, but rather demonstrating the functionality of the continuous dynamical model as a viable image classifier. In Fig. \ref{inputtarget5} the bottom row display the target pattern associated to each letter. Following the dictates of the proposed framework, the (arbitrary) patterns that define the attractors are composed by assigning to each pixel  value $\pm a$, these latter identifying the location of the two wells of the non linear reaction term. The middle row of Fig. \ref{inputtarget5} presents the noisy version of the letters.

\begin{figure}
    \centering
    \includegraphics[width=0.8\textwidth]{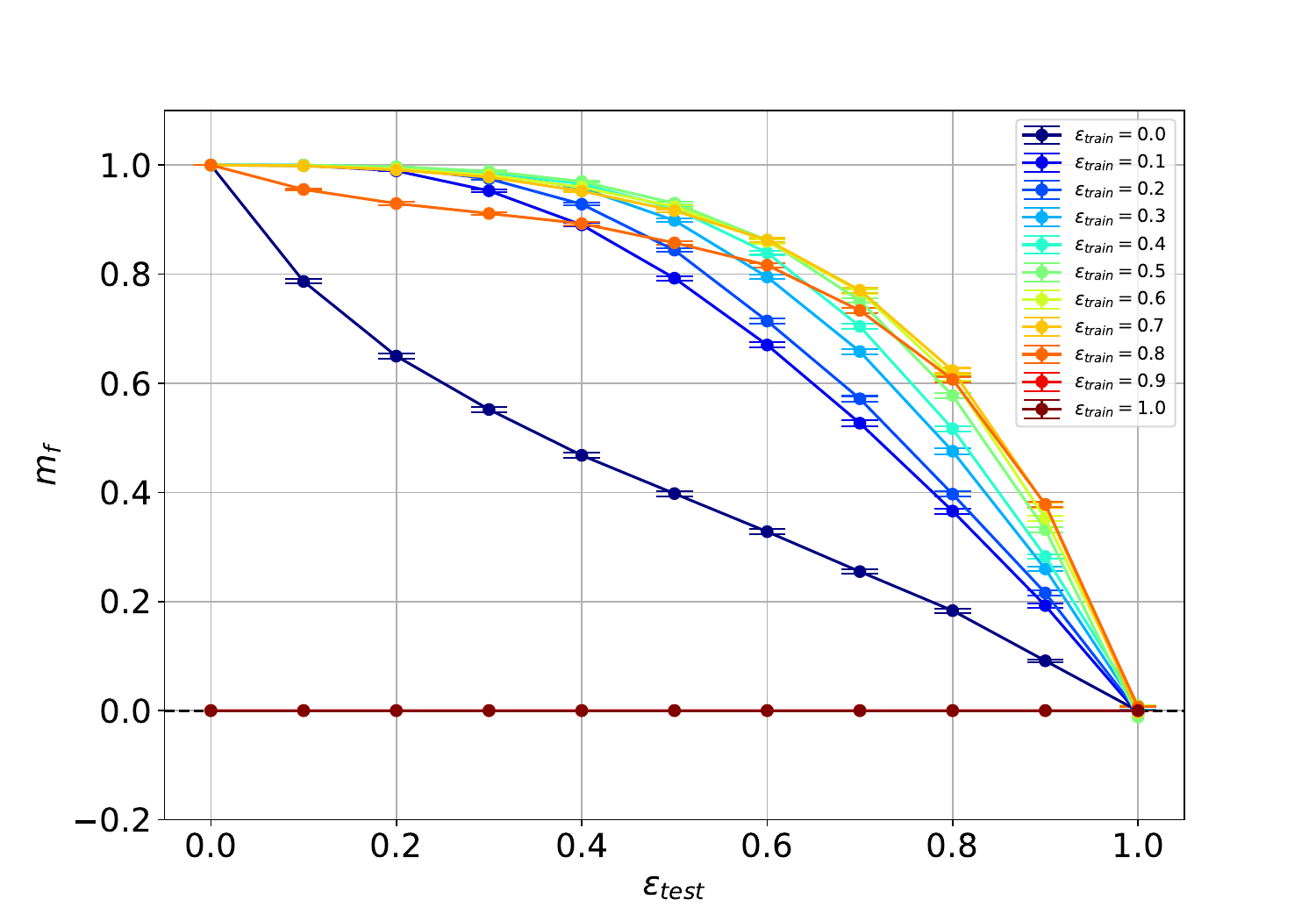}
    \caption{The figure displays the relationship between $m_f$ and the testing noise $\epsilon_{test}$ for various levels of training noise, for 5 classes. Each line represents a different training noise level, ranging from $0.0$ to $1.0$, with a corresponding color code. The horizontal line illustrates a reference point. For this analysis we set $T=40.0$, $\Delta t=0.1$.}
    \label{combinedplot5classi}
\end{figure}

As an initial analysis, we focused on the scenario where the model was trained on noise-free images ($\epsilon_{train} = 0.0$). It should be remarked that this defines a rather academic condition, where train and test sets are made of identical reservoirs, populated with multiple copies of the same (no noise) letters. As such, and for this preliminary case study, we are not really probing the ability of the system to generalize beyond the limited set of images processed under training. The detailed outcomes of this investigation are delineated in Fig.  \ref{combinedplot5classi}, blue line. The overlap between the final state and the target, i.e., the metric $m_f$, reaches an impeccable score of $1.0$. However, a discernible decrease in performance is observed as the noise level in the test images incrementally rises. The downgrade in the recorded performance is monotonous with the increase of $\epsilon_{test}$ and yields an an overlap of zero at
$\epsilon_{test} = 1.0$.

We then transitioned our focus to assess the model's prowess when trained with images bearing a noise level of $\epsilon_{train}\neq 0$.  The Fig. \ref{combinedplot5classi} offers a multidimensional perspective on the model's performance, by comparing various levels of noise during both training and testing. With a modest training noise of $\epsilon_{train}=0.1$, the model still maintains near-ideal precision for low-noise testing, but the decline in performance manifests more gradually when compared to previous observations. As we escalate the training noise, these trends persist, up to a given threshold in $\epsilon_{train}$, above which the images are too corrupted to be correctly classified.

\begin{figure}[h!]
    \centering
    \begin{subfigure}[b]{0.49\textwidth}
        \includegraphics[width=\textwidth]{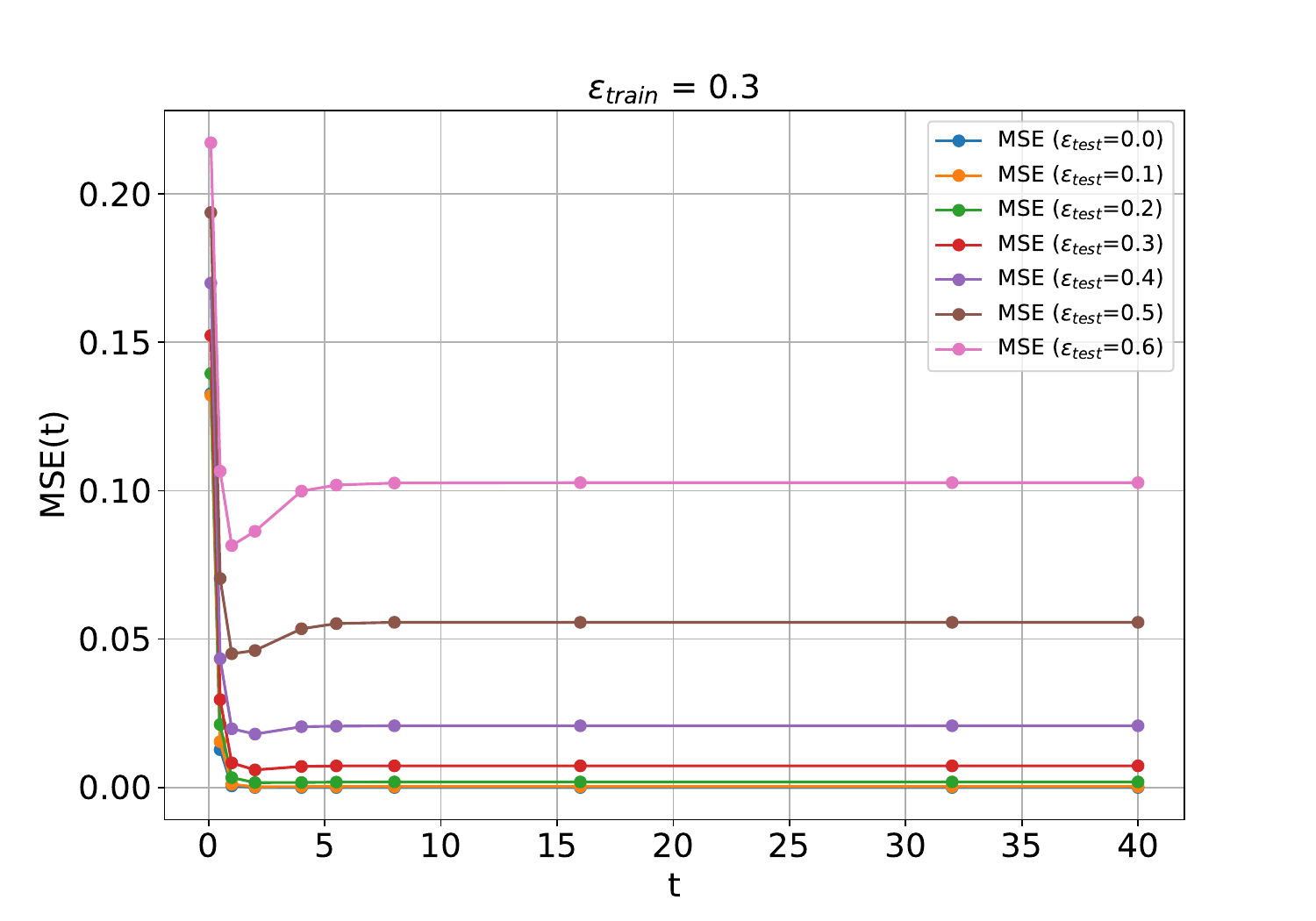}
        \caption{}
        \label{fig:MSE-A-train0.3}
    \end{subfigure}
    \hfill
    \begin{subfigure}[b]{0.46\textwidth}
        \includegraphics[width=\textwidth]{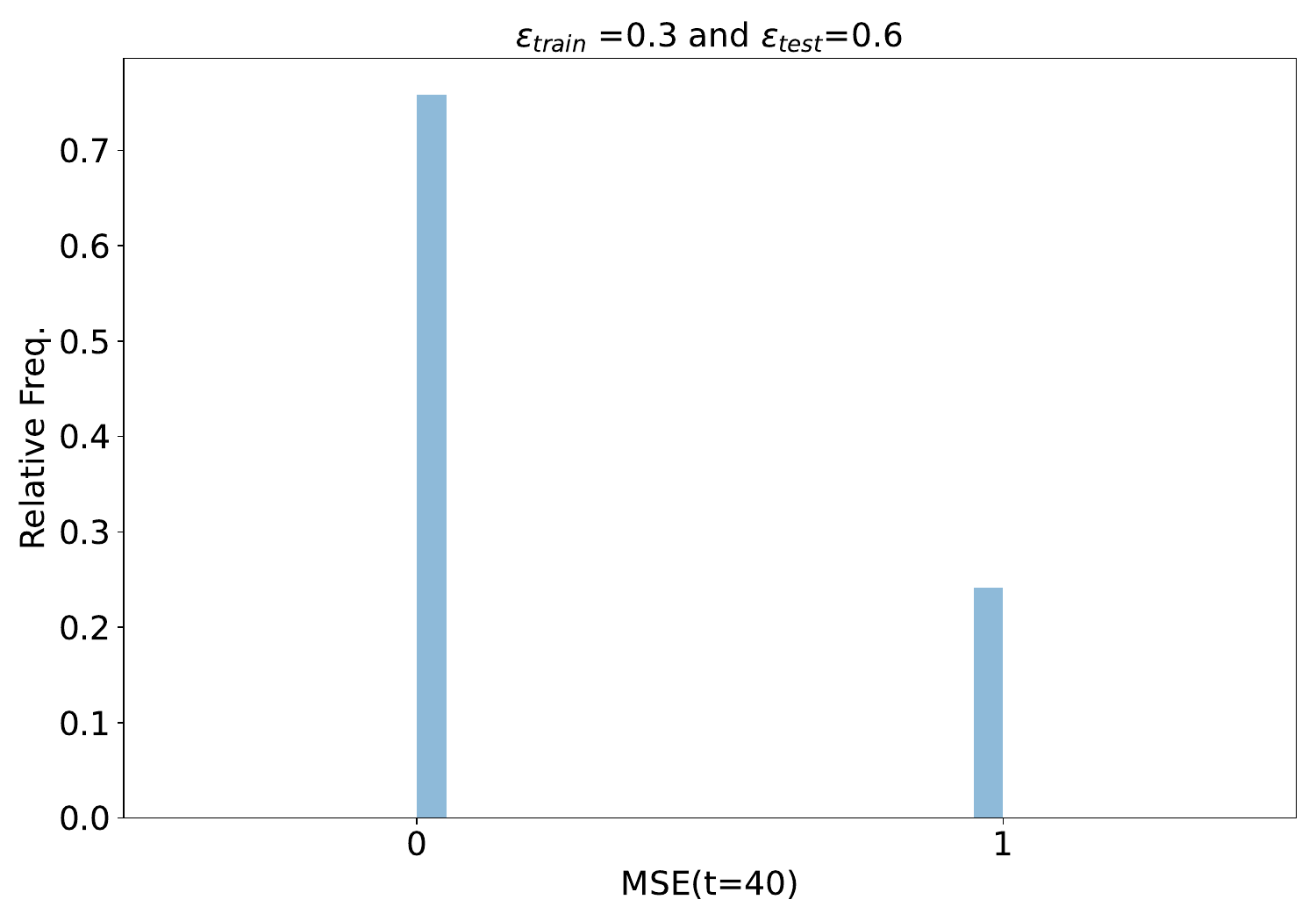}
        \caption{}
        \label{fig:histo5classi0.3}
    \end{subfigure}
    \caption{\textbf{Panel (a)}:  the temporal evolution of the Mean Squared Error (MSE) is displayed, across a sample of $2000$ images 'D', for various $\epsilon_{test}$ values, at a fixed $\epsilon_{train} = 0.3$. For scenarios where $\epsilon_{test} < \epsilon_{train}$, the MSE tends towards zero as time $t$ approaches infinity. Conversely, when $\epsilon_{test} \geq \epsilon_{train}$, the temporal estimate of the MSE increases with added noise. For this analysis we set  $\Delta t=0.1$. \textbf{Panel (b)}: the histogram of the MSE associated to  different 
    trajectories for $\epsilon_{train}=0.3$ and $\epsilon_{test}=0.6$. A significant fraction of the supplied images are correctly classified.  
     For visual clarity, all non-zero values are designated as 1.}
    \label{fig:main_MSE_acc}
\end{figure}

In Fig. \ref{fig:main_MSE_acc} \textbf{Panel (a)}, we report the trend displayed by the MSE over time $t$ for the model with $\epsilon_{train} = 0.3$. The MSE is evaluated on various test sets, identified by the level of noise introduced in the test set, namely $\epsilon_{test}$. As a first remarkable observation, we notice that the MSE reaches a stable asymptotic plateau which indirectly points to the convergence of the underlying dynamical model. The ability of the  system to classify a given datum supplied as an input, is stably maintained across time and not just referred to the limited horizon of the training time.  Further, for $\epsilon_{test} \leq 0.3$, the MSE decreases approaching zero exponentially fast. Conversely, for  $\epsilon_{test} \geq 0.4$,  the MSE stabilizes at non-zero values: trajectories within the phase space do not align with the anticipated attractors.  A detailed analysis reveals instead that, for  $\epsilon_{test} \geq 0.4$, the recorded MSE displays 
a bimodal distribution, see \ref{fig:main_MSE_acc} \textbf{Panel (b)}, with a significant fraction of images that are indeed correctly classified, also for noise amount considerably larger that the one imposed during training.
From a more fundamental point of view, the displayed bimodal distribution is reminiscent of a metastable phase associated to a first order phase transition, that putatively separates the regime where SA-nODE works properly from that where predictions are found to be inaccurate.

\begin{figure}[h!]
    \centering
    \includegraphics[width=\textwidth]{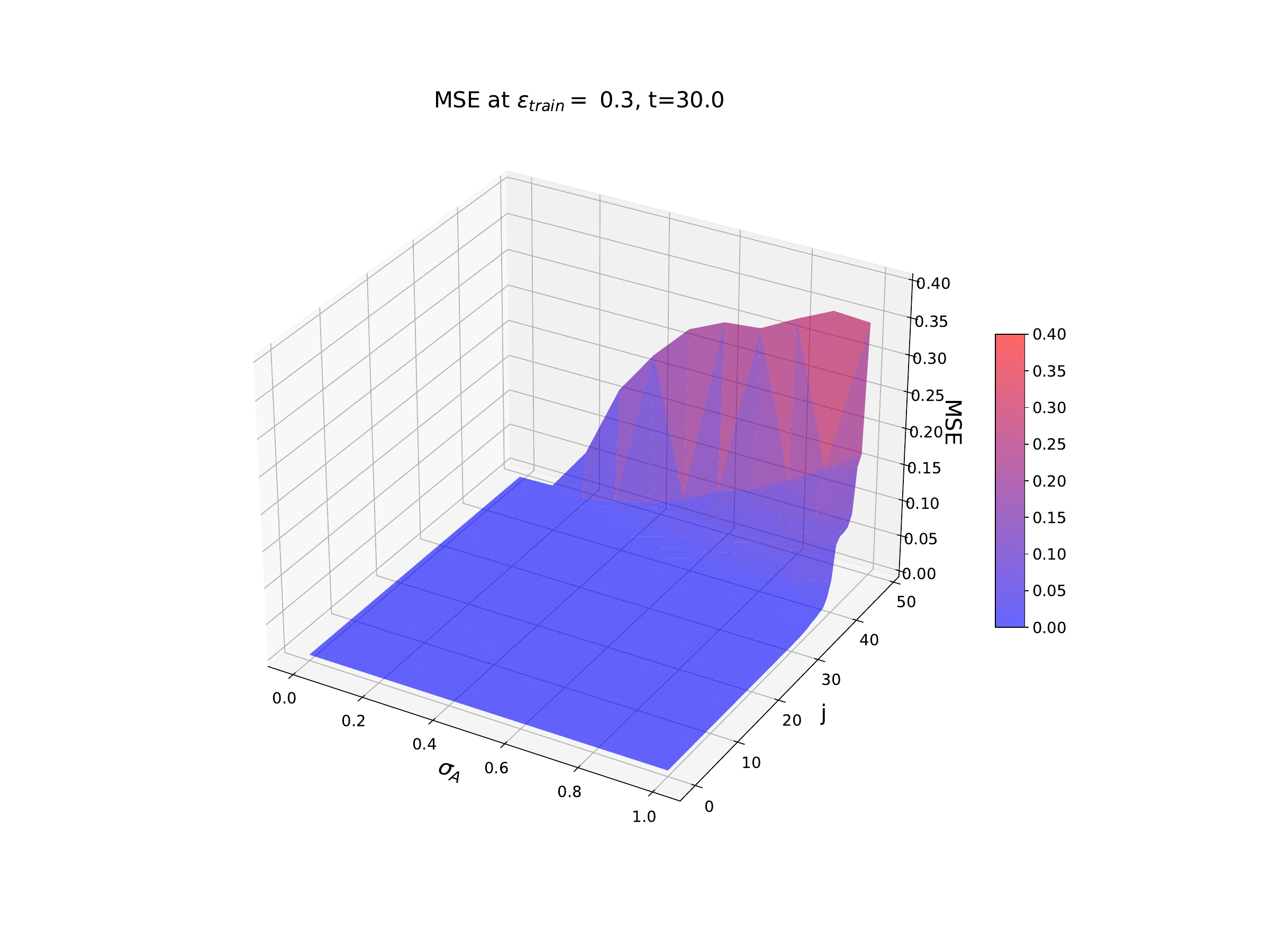}
    \caption{In the three-dimensional figure presented, the $x$-axis delineates the variance, $\sigma_{A}$, of the Gaussian perturbation applied to a specific component of the coefficient vector $\vec{c}^{(A)}$ corresponding to the $A$ letter, the $y$-axis displays the index $j$ of the perturbed component from the vector $\vec{c}^{(A)}$, and the $z$-axis represents the sample Mean Squared Error (MSE) value. Notice that the original image can be written as $\vec{x}^{(A)} = \Phi \vec{c}^{(A)}$, where $\Phi$ signifies the transformation matrix. Consequently, by leveraging the inverse of $\Phi$, we can derive the coefficient vector $\vec{c}^{(A)}$. In this representation, we introduce a perturbation exclusively to one component of the vector $\vec{c}^{(A)}$ at a given instance and compute the corresponding MSE. The indices on the $y$-axis are systematically arranged based on the ascending magnitude of the MSE, thereby illustrating the differential impact of perturbations across various components of $\vec{c}^{(A)}$ on the overall error in the reconstructed image. }
    \label{fig:unapertMSE0.3-3D}
\end{figure}

Upon training, one obtains a basis of the scanned multi-dimensional space which is tailored to the problem at hand. The basis is formed by the column vectors of matrix $ \Phi $.  Every supplied item $\vec{\overline{x}}$ can be decomposed by using the aforementioned basis. The set of obtained coefficients, $\vec{\overline{c}}=\Phi^{(-1)} \vec{\overline{x}}$ returns a complete and equivalent representation of the analyzed datum $\vec{\overline{x}}$. This  framework can be used to probe  the model's robustness and vulnerability to external source of disturbance from a  different perspective, alternative to perturbing each individual pixel (which amounts to operate with the canonical basis viewpoint). More specifically, we can proceed by perturbing individual eigen-directions - or punctually alter each coefficient $\vec{\overline{c}}$ of the above expansion. In doing so, several pixels get simultaneously modulated by the external noise source, following a pattern of correlated activation that indirectly stems from the accomplished training. In doing so we empirically, senses the size of the basins of attraction, post training, by using a correlated version of noise that is self-consistently shaped by the learning protocol. 

 Fig. \ref{fig:unapertMSE0.3-3D} reports on this analysis, where the robustness of each eigen-direction to a Gaussian perturbation $ \mathcal{N}(0,\sigma)$ is probed. The MSE, calculated in a steady state, serves as a benchmark metric. We observe that, for a model trained with $ \epsilon_{train}=0.3 $, many coefficients are resilient to individual perturbations. Stated different only few directions can trigger the system unstable. In fact, the vast majority of imposed collective eigen-perturbations do not affect the ability of the system to carry out the assigned classification task. In a world where images are frequently subjected to noise and interference, understanding noise resistance through coefficient perturbation can guide the implementation of resilient designs and/or information compression strategies, ensuring that essential information is preserved despite reductions.

\begin{figure}[h!]
    \centering
    \includegraphics[width=0.7\textwidth]{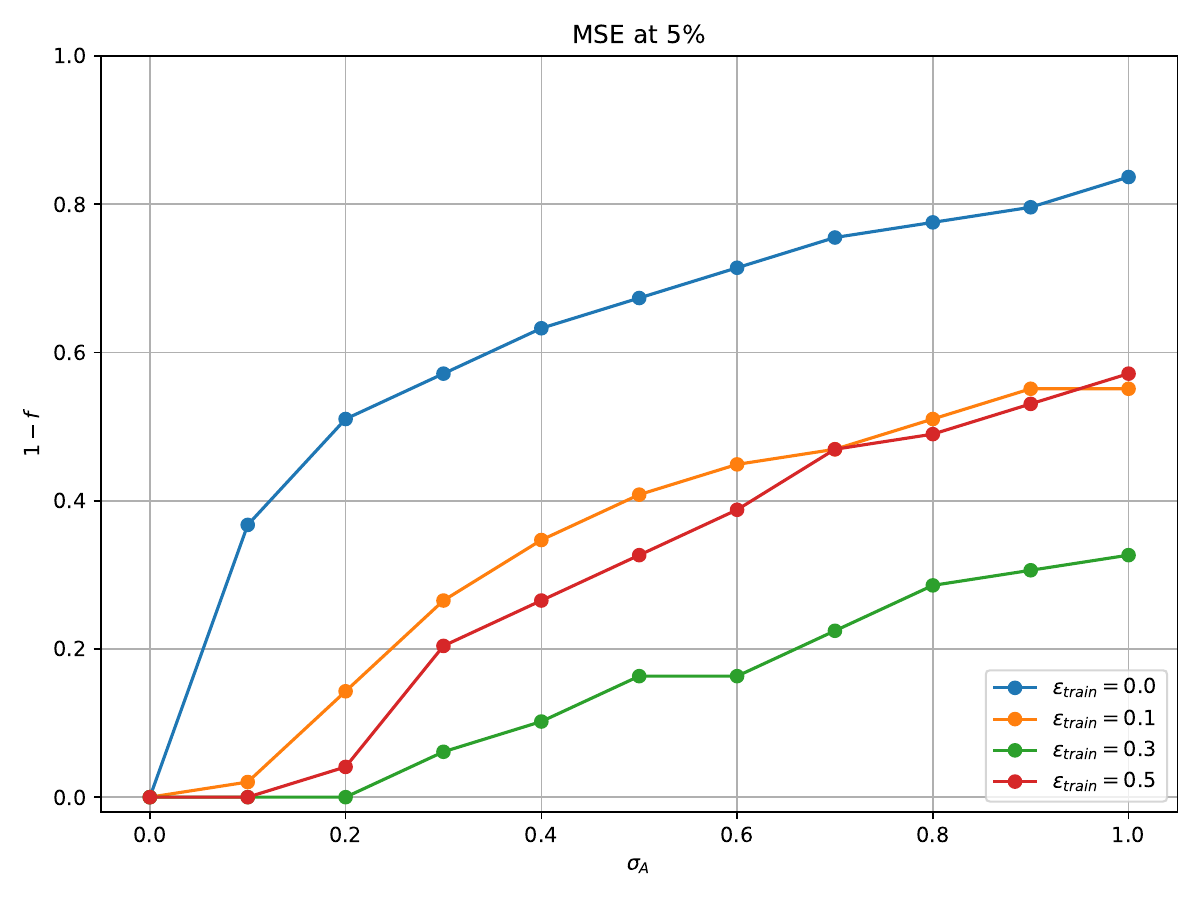}
    \caption{Analysis of model robustness by examining the fraction \(f\) of coefficients that can be simultaneously perturbed with Gaussian noise (\(\sigma_A\)) without exceeding a 5\% degradation in MSE. $1-f$, conversely, is related to the number of sensitive eigen-directions (the smaller the better). The primary objective is to ascertain the threshold beyond which the cumulative effect of these perturbations impedes the accurate classification of a reconstructed image (here letter $A$). This analysis is carried out for different choices of $\epsilon_{train}$. A discernible observation from the figure suggests the existence of an optimal noise level up to which the perturbations remain inconsequential for the classification task. In this specific instance, the optimal value is identified as $\epsilon_{train}=0.3$.}
    \label{fig:MSE-many-pert-coef}
\end{figure}

Fig. \ref{fig:MSE-many-pert-coef} shows how the addition of noise in the training process results in a marked increase in the model's robustness, which can be assessed in reciprocal space. Specifically, we reason as follows: let us focus on the letter A. The coefficients in the reciprocal space of this data can be perturbed with a Gaussian distribution of variance \(\sigma_A\), causing a degradation of the MSE. As shown in Fig. \ref{fig:unapertMSE0.3-3D}, several modes are, in principle, robust to perturbations applied to individual coefficients. However, from this analysis it is not obvious to anticipate what could  happen when a fraction \(f\) of the coefficients gets simultaneously perturbed with Gaussian random noise of variance \(\sigma_A\) and how this collective perturbation eventually reverberates on the recorded MSE.

To understand the effect of training with noise \(\epsilon_{Train}\), we fix the degradation of the MSE at 5\% and the perturbation amplitude \(\sigma_A\). A natural question hence arises: how many coefficients can be perturbed without exceeding the 5\% degradation threshold in the MSE? Alternatively, what fraction (1-\(f\)) of the eigenvectors must not be perturbed? This latter fraction is, in our view, a measure of the network's robustness. Indeed, if it is low, only a few coefficients need to remain unperturbed, while the remaining \(f\) can be perturbed simultaneously without causing a degradation greater than 5\% of the MSE. The result of this analysis is reported in Fig.6 against different training-noise amplitudes. Such representation shows the existence of an optimal training-noise ($\epsilon_{Train} =0.3$) level. With this latter choice, almost 70 \% of the eigenvectors ca be perturbed with a maximal strength of the imposed noise ($\sigma_A=1.0$) without observing a relative difference in the MSE larger than 5 \%. The fraction of relevant eigenvectors is, moreover, consistently lower with respect to the other training-noises.

Upon completing the analysis of the synthetic dataset, we can now shift our focus to assess the robustness of Sa-nODE in comparison to the most commonly cited benchmarks in literature: MNIST and Fashion MNIST.

\section{Applying SA-nODE to MNIST and Fashion MNIST}\label{sec:MNIST}

In the preceding section we introduced the concept of SA-nODE and applied it to the analysis of a synthetic dataset, with different grades of imposed noise. To further test the adequacy of the proposed scheme, we turn to considering the so called MNIST and Fashion MNIST datasets.

The MNIST (Modified National Institute of Standards and Technology) dataset \cite{deng2012mnist} is one of the most iconic and widely utilized datasets in the field of machine learning, particularly in the domain of pattern recognition and computer vision. Stemming from a rich history, it has played a pivotal role as a benchmark dataset, testing a plethora of algorithms and machine learning techniques over the years. 

The MNIST dataset consists of a collection of $70000$ handwritten digit images. These gray scale images are uniformly sized at $28\times28$ pixels (i.e., $N=784$), with digits centered in the frame. The dataset is subdivided into a training set of $60000$ images and a test set comprising $10000$ images. Each image is labelled with the corresponding digit it represents, ranging from $0$ to $9$.

The simplicity of the dataset, combined with its relatively small size, makes it suitable for beginners to explore the intricacies of various machine learning algorithms without the need for substantial computational resources. Furthermore, the dataset's balanced composition, with an almost equal number of samples for each digit, ensures that models trained on it are not biased towards any particular digit.

The accuracy of the trained SA-nODE algorithm can be computed from evaluation of $m_f$, as introduced above. More specifically, we assumed that an image is correctly classified only if the similarity $m_f^{(j)}= \frac{1}{N}\sum_{i=1}^N sign(x^{ (j)}_i)sign(\overline{\phi}^{(y^{(j)})}_i)$ is greater than or equal to $95\%$. 

\begin{figure*}
    \centering
    \includegraphics[width=\textwidth]{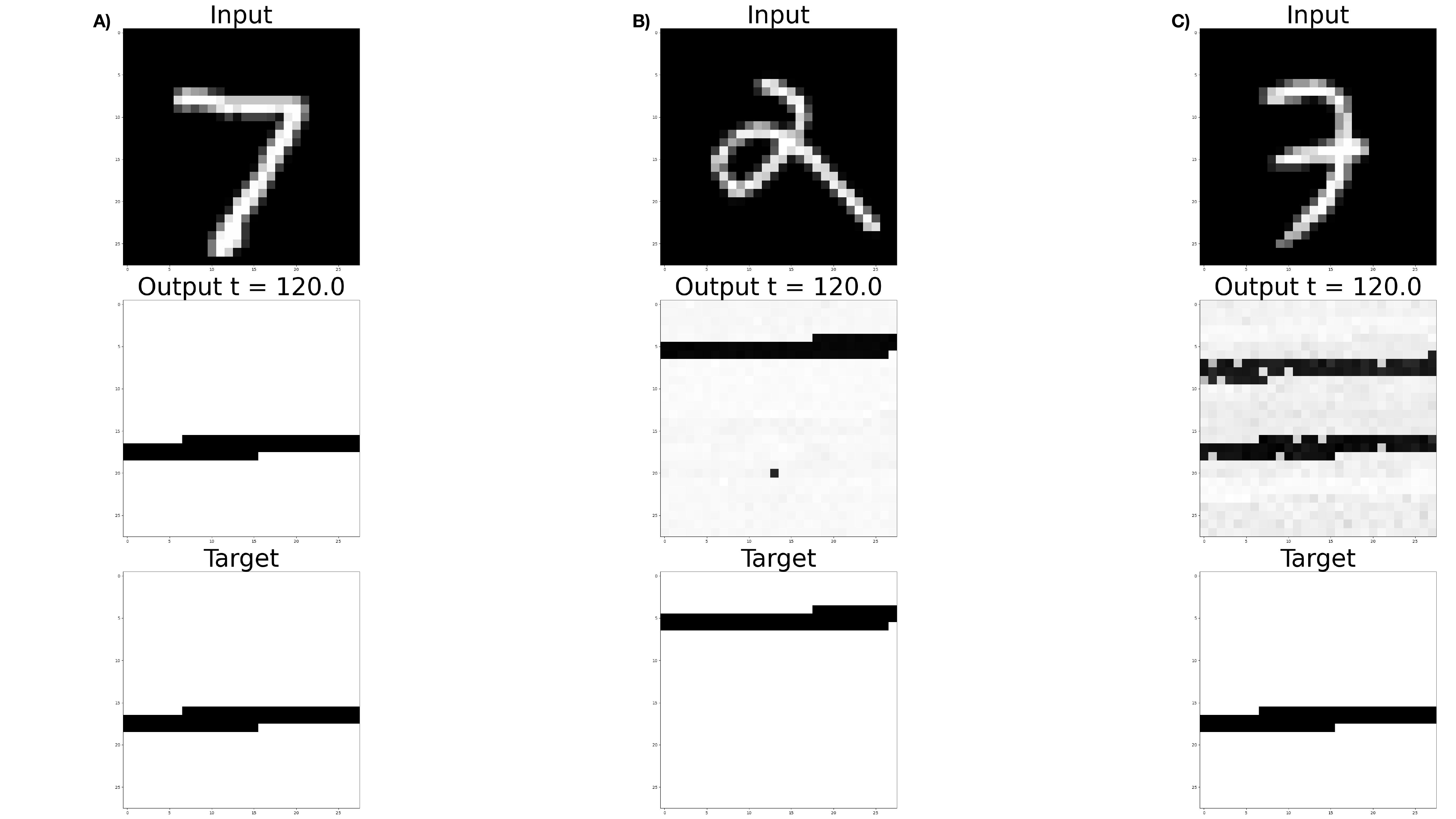}
    \caption{Illustration of three scenarios where the model classifies images. \textbf{A)} This scenario represents the epitome of accurate classification, where $m_f^{(j)} = 1$. It showcases flawless categorization with no corrupted pixels in the output targets. \textbf{B)}In this case, $0.95 \leq m_f^{(j)} < 1$, demonstrating targets with minimal pixel differences. Despite these slight variations, the model maintains accurate classification, highlighting its robustness. \textbf{C)} Indicates potential misclassification or the presence of overlapping attractors, where $m_f^{(j)} < 0.95$. For this analysis we set $T=120.1$,  $\Delta t=0.1$.}
    \label{Figure_MNIST_class}
\end{figure*}

In formulae, the accuracy $(\mathcal{H})$ of our algorithm is formulated as the average over the test dataset ($\mathcal{D}_{test}$):

\begin{equation}
\mathcal{H} = \frac{1}{|\mathcal{D}_{test}|} \sum_{j=1}^{|\mathcal{D}_{test}|} \delta(m_f^{(j)}),
\end{equation}
where
\begin{equation}
   \delta(m_f^{(j)})= \begin{cases} 1 & \text{if } m_f^{(j)} \geq 0.95 \\ 0 & \text{otherwise} \end{cases}
\end{equation}

This allows us to account for images that exhibit minimal pixel differences from the correct attractors, while excluding those with more than $5\%$ of corrupted pixels. This refined  metric provides a comprehensive evaluation of our algorithm's classification capabilities, in terms of reported accuracy.

In such conditions, the accuracy achieved by SA-nODE is $98.06\%$. Among the $10^4$ test images, $9776$, an overwhelming majority, are perfectly classified with $m_f^{(j)} = 1$, indicating that no pixels are corrupted in the output images. In other words, the $9776$ initial conditions have converged to the attractors within a specified maximum time frame and the state achieved corresponds accurately with the classification requirements. Furthermore, $30$ images exhibit $0.95 \leq m_f^{(j)} < 1$, and the remaining images, though a small fraction, have $m_f^{(j)} < 0.95$. 

To provide a comprehensive visual insight into the varying degrees of pixel-wise similarity, three cases typical examples are provided in 
Fig. \ref{Figure_MNIST_class}: Case A shows the perfect classification, thus $m_f^{(j)} = 1$,  with no corrupted pixels. Case B displays an instance in the range $0.95 \leq m_f^{(j)} < 1$, showcasing output images with minimal pixel differences, while maintaining accurate classification. Case C indicates potential misclassification, without necessarily having corrupted pixels or overlapping attractors. In this particular instance, we can observe the algorithm's uncertainty as it produces an output that is a combination of the targets for both a three and a seven.

As a second application, we consider Fashion MNIST dataset. This is specifically designed as a drop-in replacement for the classical MNIST dataset, albeit with a contemporary twist. Instead of handwritten digits, Fashion MNIST encompasses a variety of images pertaining to clothing and fashion articles. Introduced by Zalando, an e-commerce company, \cite{xiao2017fashion} the dataset responds to the criticism that MNIST was considered too easy, or even overused, in the deep learning community.
While the images in the Fashion MNIST might appear simple, the subtle nuances between certain clothing categories make classification a more challenging task than distinguishing between handwritten digits.

Fashion MNIST consists of $70000$ images, representative of $10$ distinct fashion categories, such as t-shirts, trousers, pullovers, dresses, coats, sandals, and more. Similar to the MNIST dataset, these gray scale images are standardized to a size of $28\times 28$ pixels. The dataset is neatly split into $60000$ training images and $10000$ test images. Each image is associated with a label that denotes one of the ten categories.

SA-nODE  achieves an accuracy of $88.21\%$. More precisely,  among the $10^4$ test images, $8740$, an overwhelming majority, are perfectly classified with $m_f^{(j)} = 1$, indicating that no pixels are corrupted in the output images. Furthermore, $81$ images exhibit $0.95 \leq m_f^{(j)} < 1$, and the remaining images, though a small fraction, have $m_f^{(j)} < 0.95$.

\section{Conclusion}\label{sec:discuss}

We have here proposed a variant of the Neural ODEs architecture, which accounts for a closed analytical expression of the imposed reaction term. Individual nodes are in particular assumed to be subject to a local reaction which follows from a prototypical double well potential. As such, the uncoupled dynamics can accommodate for two symmetric equilibria. These latter solutions define a minimal alphabet that we use to construct a set of arbitrarily crafted attractors of the coupled (spatially extended) system dynamics. The coupling among computing nodes, each associated to a pixel of the image to be classified, is linear and topologically encoded in an adjacency weighted matrix, whose elements represent the target of the training process. This latter matrix is formulated in reciprocal domain and the target attractors are assigned to belong to its kernel. Stationary stability is a priori granted by forcing the trained eigenvalues to populate a given domain, as stipulated by a linear stability analysis. Classifying amounts to shaping the basin of the planted attractors, so that  different items will be eventually directed towards distinct targets, depending on their specific category of pertinence. In essence, SA-nODE, as we chose to denote the algorithm, is made of a finite set of particles or spins, each of which jumping between two allowed states, identified by the positions of the symmetric potential wells.  Because of mutual (linear) interactions, and following spectral training, the particles are eventually frozen in a heterogeneous (in the space of the pixel) and stationary stable pattern, yielding an image that is uniquely representative of the processed element. The proposed classification strategy has been successfully tested against mock and reference datasets, as reported in the main body of the paper. The interest of SA-nODE resides in having constructed a simple, dynamically sound model, which yield competitive classification scores. The model takes a closed, compact and intuitive form, which can be further dissected by leveraging on the vast  arsenal of non-linear dynamics tools, to shed light, from a different angle, on the inherent ability of neural networks to effectively learn and decide. We anticipate that other models, including several  of biomimetic inspiration, thus relevant to computational neuroscience, could be proposed, which yield dynamically assisted classification via stable attractors. We will report in future works on these possible extensions.

\section*{Acknowledgments}

This work is supported by \#NEXTGENERATIONEU (NGEU) and funded by the Ministry of University and Research (MUR), National Recovery and Resilience Plan (NRRP), project MNESYS (PE0000006) "A Multiscale integrated approach to the study of the nervous system in health and disease" (DR. 1553 11.10.2022).

\bibliography{ref}

\appendix

\section{Converge to the continuous model} \label{app::continuity}

\begin{figure}[htbp]
    \includegraphics[width=\textwidth]{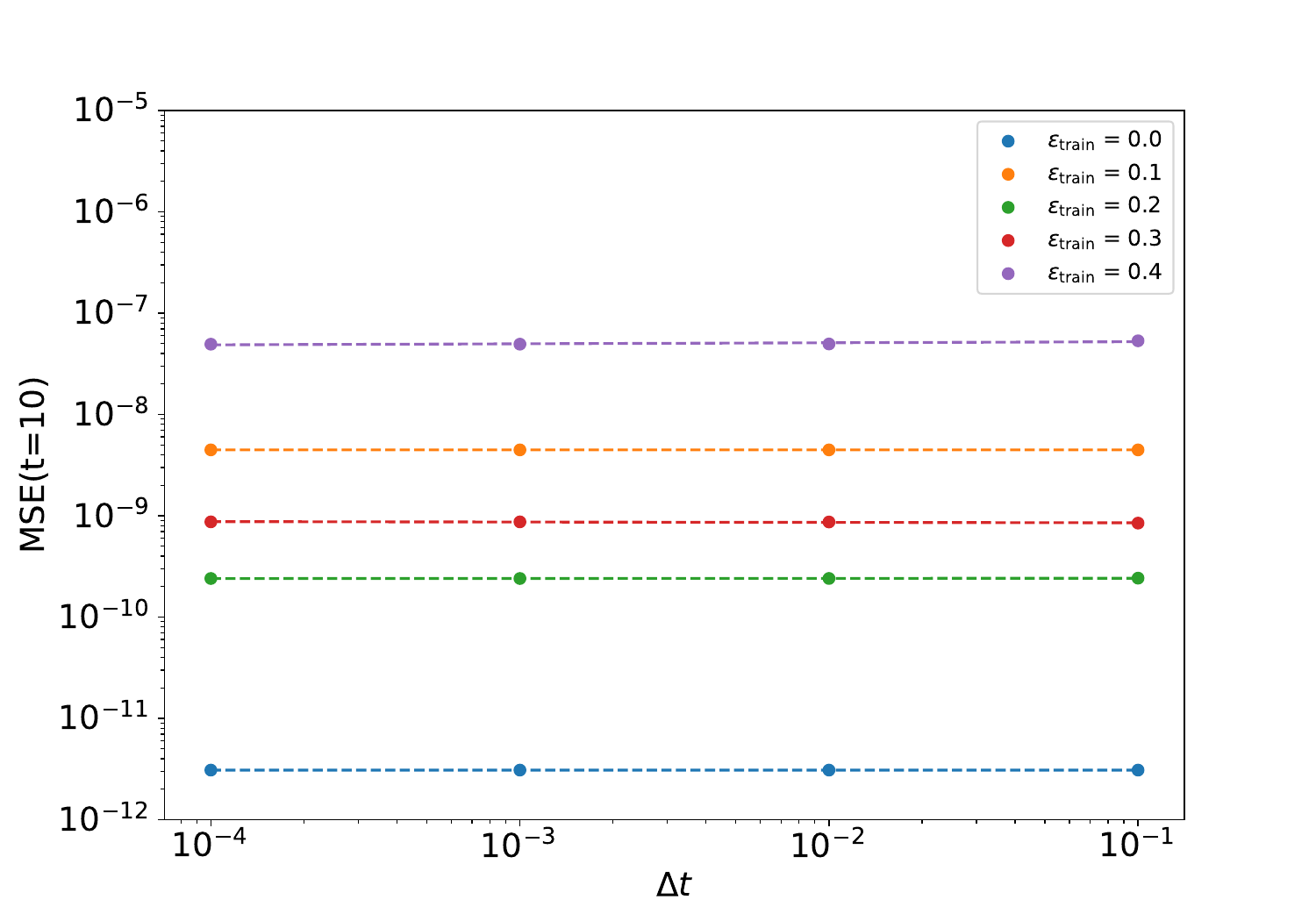}
    \caption{The plot showcases the MSE at $t=10$ as a function of the chosen time step $\Delta t$ for numerical integration, employing the fourth-order Runge-Kutta method.}
    \label{figcontinuity}
\end{figure}

To challenge numerical convergence, we varied the integration timesteps, $\Delta t$, and recovered the results reported in Fig. \ref{figcontinuity}. We can hence utterly conclude that the reported analysis provides a faithful representation of the underlying continuum dynamical system.  These numerical integrations were carried out using the fourth-order Runge-Kutta method \cite{butcher2016numerical}. This method is employed for its precision, having a truncation error of $O(\Delta t^4)$, where $\Delta t$ is the time step of the integration.

\section{Invertibility of the SA-nODE algorithm} \label{app::invertibility}

Invertible machine learning models can be used for data compression, enabling high-fidelity data decompression or synthetic data generation.  Invertibility also aids in model interpretability by maintaining information throughout the network's layers, which can illuminate the model's decision processes. With these applications in mind we set we begin with an image, represented as $\vec{x}(t=0)$, serving as the initial condition of our dynamics. This image is allowed to evolve in accordance with its inherent dynamical laws. Post-training, it is constructively known that the image concludes its dynamical trajectory at an attractor.

A pivotal question arises at this juncture: Having the output of our classification algorithm ($x(T)$), is it possible to reconstruct the originating image ($x(0)$)?

\begin{figure}
    \centering
    \includegraphics[width=9cm]{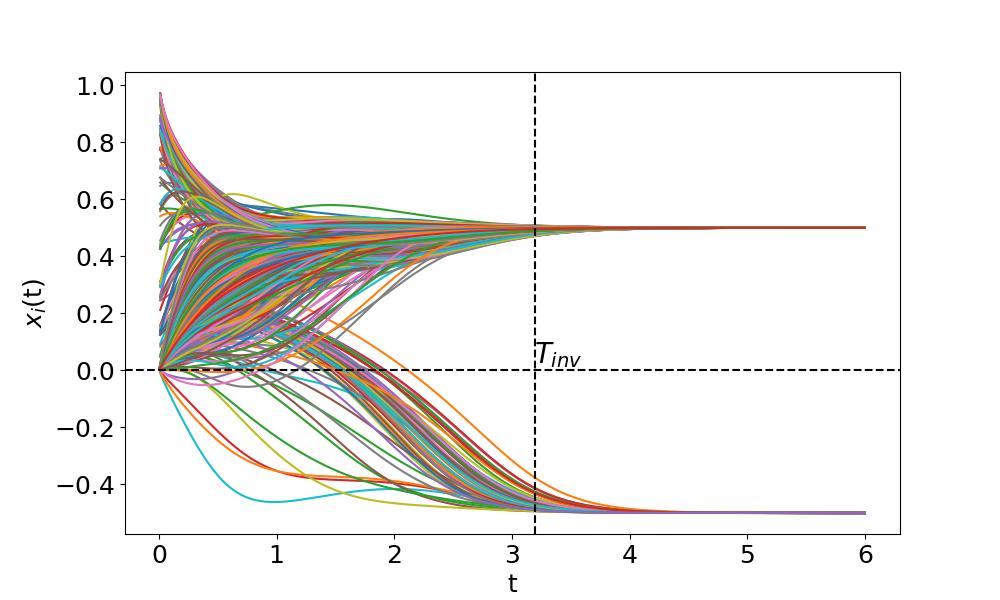}
    \caption{The figure depicts the temporal evolution of the $784$ pixels of an MNIST image corresponding to the digit $0$. Notably, all curves converge to distinct clusters in the positive and negative planes once $t > T_{inv}=3.2$, where $T_{inv}$ stands for the time where the inversion takes place. For this analysis we set  $\Delta t=0.01$.}
    \label{ExampledynamicsMNIST}
\end{figure}

Addressing this question, we find that SA-nODE possesses a mechanism for reversing its dynamics. By employing a straightforward transformation, $\tau=T-t$ \cite{han2018solving, marino2016advective}, which effectively reverses the dynamics of the system, we can integrate the dynamics backwards and recapture the initial image. The modified evolutionary law for this backward integration becomes $\vec{\dot{x}}=-\vec{F}(\vec{x})$, with the derivative now taken concerning the variable $\tau$. It is noteworthy to mention that the choice of $T$ in the transformation $\tau=T-t$ is crucial. $T$ should not be excessively large, but large enough to ensure that the system has reached the attractor, as illustrated in Fig. \ref{ExampledynamicsMNIST}.

\begin{figure}
    \centering
    \includegraphics[width=9cm]{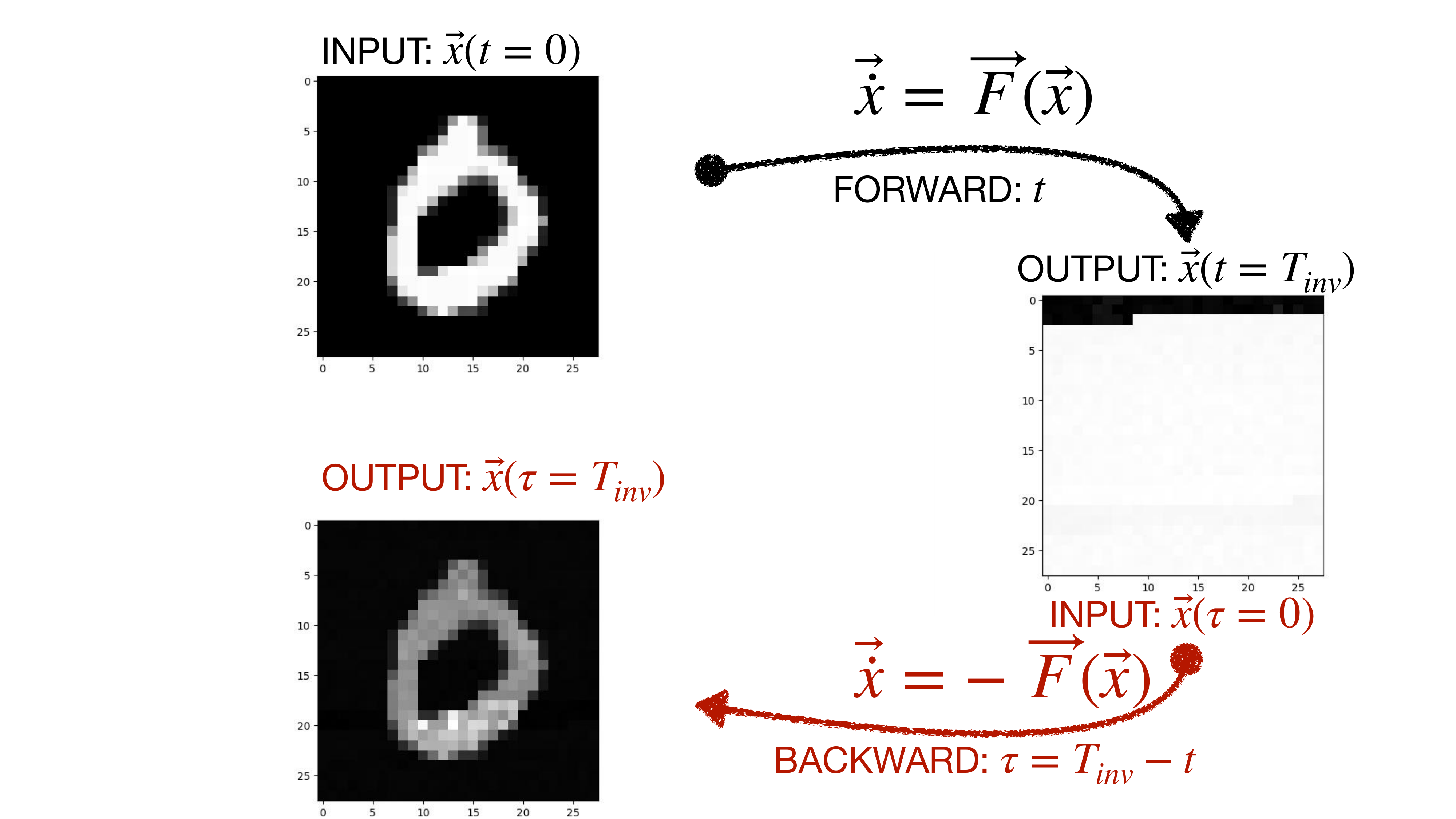}
    \caption{The figure illustrates the process of reversing the time arrow for an MNIST image corresponding to the digit $0$. The classification can be easily verified by the image on the right, while the recovered image can be seen at the bottom on the left of the figure. For this analysis we set  $\Delta t=0.0001$, $T_{inv}=3.2$.}
    \label{MNISTinv}
\end{figure}

\begin{figure}
    \centering
    \includegraphics[width=9cm]{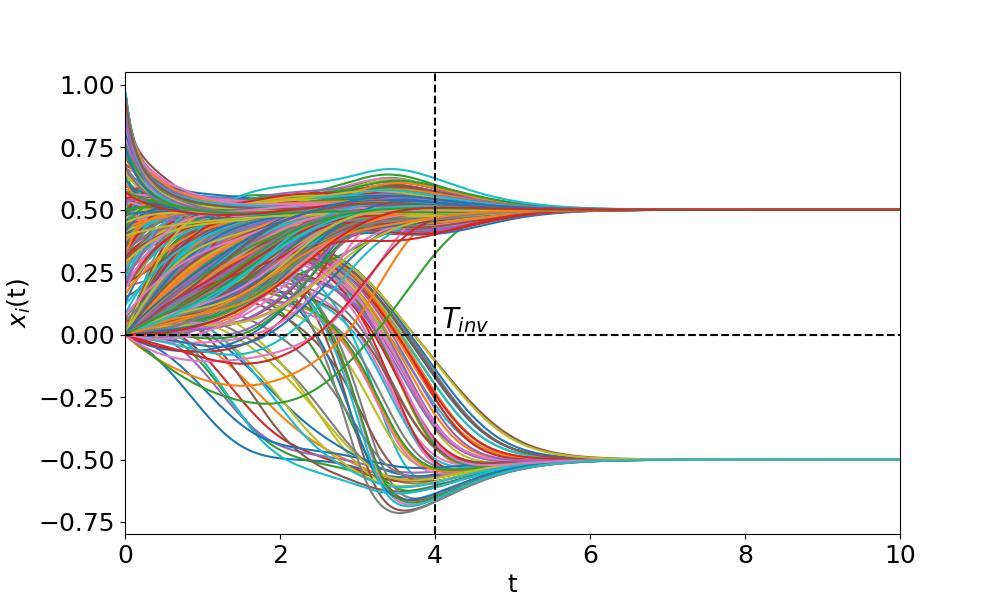}
    \caption{The figure depicts the temporal evolution of the $784$ pixels of an Fashion MNIST image corresponding to trousers in Fig. \ref{FMNISTinv}. Notably, all curves converge to distinct clusters in the positive and negative planes once $t > T_{inv}=4.0$. For this analysis we set  $\Delta t=0.01$.}
    \label{ExampledynamicsFMNIST}
\end{figure}

\begin{figure}
    \centering
    \includegraphics[width=9cm]{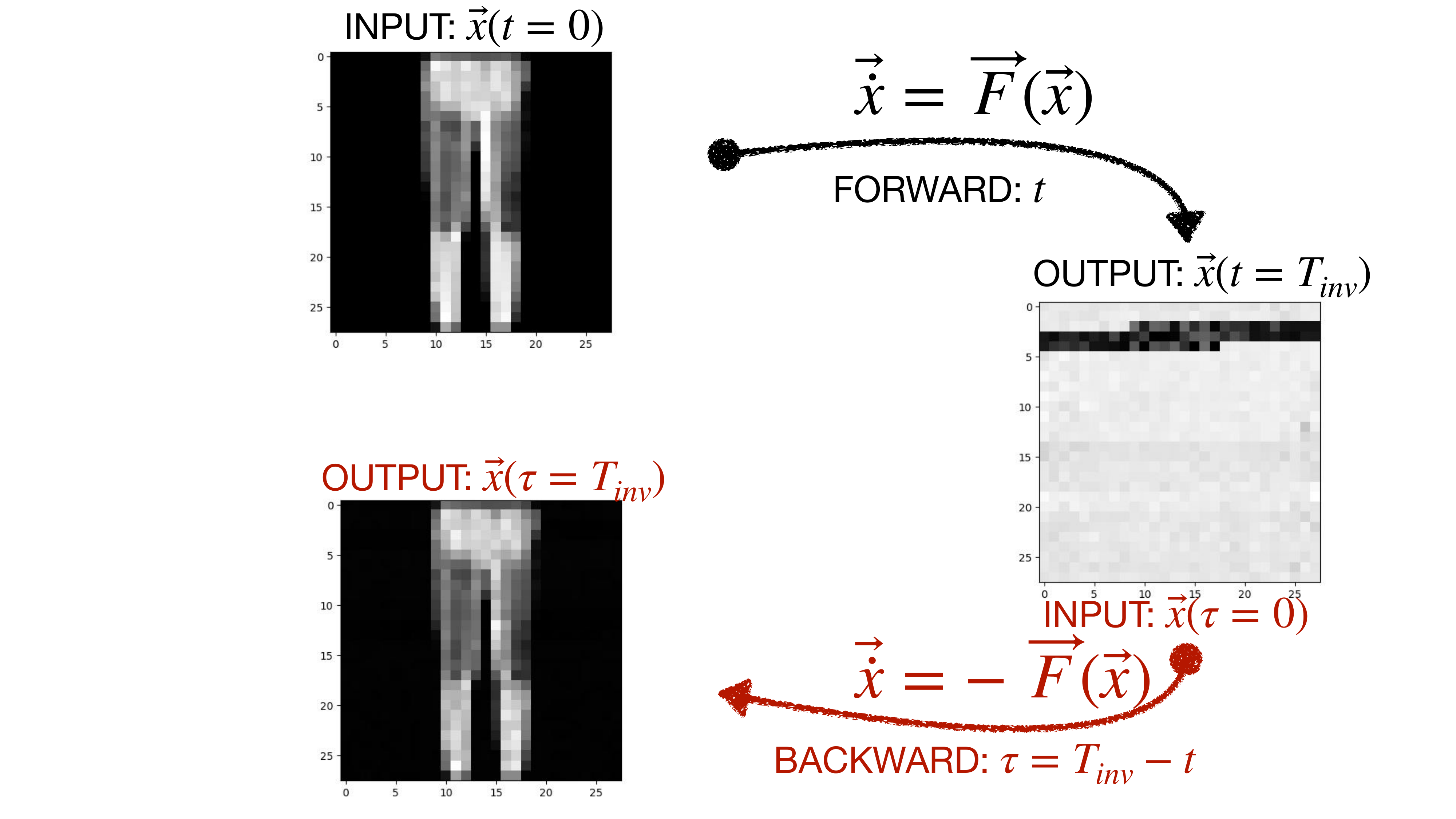}
    \caption{The figure illustrates the process of reversing the time arrow for Fashion MNIST image. The classification can be easily verified by the image on the right, while the recovered image can be seen at the bottom on the left of the figure. For this analysis we set  $\Delta t=0.00001$, $T_{inv}=4.0$.}
    \label{FMNISTinv}
\end{figure}

Figure \ref{ExampledynamicsMNIST} and  \ref{MNISTinv} (resp. \ref{ExampledynamicsFMNIST} and \ref{FMNISTinv}) provide a graphical representation of the journey from a starting image to the recovered image, obtained via the backward dynamic integration. Notably, the figures obtained closely resembles the input image, with any discernible differences likely attributed to numerical errors inherent in the integration process.

\end{document}